\documentclass[aip, pof, preprint, fleqn]{revtex4-1}
\usepackage{graphicx}
\usepackage[utf8]{inputenc} 
\usepackage{amsmath} 
\usepackage{amsthm}
\usepackage{enumerate}
\usepackage{epstopdf}

\begin{document}

\title{Optimal control of circular cylinder wakes using long control horizons}

\author{Thibault L. B. Flinois}
\email{t.flinois11@imperial.ac.uk}
\affiliation{Department of Aeronautics, Imperial College London, London SW7 2AZ, UK}
\author{Tim Colonius} 
\affiliation{Department of Mechanical and Civil Engineering, California Institute of Technology, Pasadena, CA 91125, USA}

\date{\today}

\begin{abstract}
The classical problem of suppressing vortex shedding in the wake of a circular cylinder by using body rotation is revisited in an adjoint-based optimal control framework. The cylinder's unsteady and fully unconstrained rotation rate is optimized at Reynolds numbers between 75 and 200 and over horizons that are longer than in previous studies, where they are typically of the order of a vortex shedding period or shorter. In the best configuration, the drag is reduced by $19\%$, the vortex shedding is effectively suppressed, and this low drag state is maintained with minimal cylinder rotation after transients. Unlike open-loop control, the optimal control is shown to maintain a specific phase relationship between the actuation and the shedding in order to stabilize the wake. A comparison is also given between the performance of optimizations for different Reynolds numbers, cost functions, and horizon lengths. It is shown that the long horizons used are necessary in order to stabilize the vortex shedding efficiently.
\end{abstract}

\maketitle

%%%

\section{Introduction} 

The flow past a circular cylinder is often used to test and compare the ability of control methods to suppress flow instabilities. When the Reynolds number of the flow ($Re=U_\infty D/\nu$, where $U_\infty$ is the inflow velocity, $D$ is the cylinder diameter and $\nu$ is the kinematic viscosity) reaches about $47$, the interaction between the two shear layers becomes linearly unstable, resulting in a supercritical Hopf bifurcation\cite{Provansal1987}. For higher Reynolds numbers, a stable limit cycle occurs in the form of the well-known Von K\'{a}rm\'{a}n vortex street, where vortices are shed in turn from each side of the body. 

The suppression of the wake instability for $Re>47$ has been thoroughly studied both for the simplicity of the cylinder geometry and for the industrial interest in reducing bluff-body form drag and wake fluctuations. A plethora of control methods have been applied to this setup and some of the most important and successful approaches were reviewed by Choi\cite{choi2008}. Usually, flow control strategies are categorized into \textit{passive} methods, which do not require an energy input -- such as splitter plates \cite{Anderson1997} or control cylinders \cite{Schumm1994,Strykowski1990} -- and \textit{active} methods, which do -- e.g. cross-flow body displacement \cite{Siegel2006} or cylinder rotation \cite{Tokumaru2006a}. 

More recently, there has been an increasing interest in {\it closed-loop} active control methods, whereby the actuation is modified in real-time, based on information from the flow-field. This allows set-point tracking, as well as disturbance rejection, and crucially, it has the potential to stabilize a system about an unstable operating point, thus only requiring a small amount of energy input after transients as shown in several studies\cite{Ahuja2010,Tu2012,Dowling2005}. 

The feedback control approach chosen here is \textit{adjoint-based optimal control}. This method consists in finding and applying the control input that locally minimizes a cost function over a given time horizon. Although this method can theoretically be implemented as a real-time feedback control strategy for some systems, this is not currently computationally feasible for fluid flows. For such systems, it can therefore be seen as an optimization technique where the control is optimized offline. A key advantage of this method is that it does not require an a priori knowledge of the physical mechanism that will minimize the cost function, so it can be used to gain intuition about unsuspected but effective control strategies. For instance Joe {\it et al.} \cite{Joea} applied optimal control to the flow over a separated flat plate with a jet located near the trailing edge and found that a phase-locked, pulse-like waveform was optimal as it interacted with the vortices shed in the wake in a more efficient and robust way than sinusoidal forcing. Based on these results, Joe {\it et al.} \cite{Joea} then designed a simple phase-locked controller using a similar waveform, leading to a near-optimal performance, without the need for expensive estimators. Additionally, the cost of finding this optimal control does not increase regardless of the number of control inputs. For instance Bewley {\it et al.}\cite{Bewley2001a} applied optimal control to re-laminarize a turbulent channel flow, where every point in the channel was used for blowing and suction. Finally, adjoint-based optimal control can be applied to steady and unsteady flows, as well as linear and nonlinear flows. Most other control approaches are not this versatile especially with regard to the nonlinearity of the Navier-Stokes equations. In many contexts, this approach can therefore be seen as providing an upper limit on the achievable performance for a given control configuration. However, because in most cases the problem is nonlinear and non-convex, many studies have found that the results can be strongly influenced by the definition of the cost function\cite{Min1999,Bewley2001a} and by the length of the optimization horizon\cite{Protas2002,Li2003,Bewley2001a}. The importance of carefully choosing the control problem definition is therefore also investigated in this article.

In the present study, rotation of the cylinder about its main axis is chosen as the control method. The experimental work of Tokumaru and Dimotakis\cite{Tokumaru2006a} demonstrated the potential for this control method to affect the flow-field and reduce the drag of a cylinder by up to roughly $80\%$ at $Re=1.5\times 10^4$ and has prompted many other authors to investigate the effect of oscillatory cylinder rotation on the development of the vortex street at different Reynolds numbers, e.g. \cite{Choi2002,Thiria2006,Shiels2001,Lu2011,Bergmann2006b}. In these studies, the cylinder's sinusoidal rotation needs to be vigorous for the control to successfully reduce drag: typically the amplitude of the tangential velocity on the cylinder surface is several times larger than the inflow velocity, and the Strouhal number associated with the cylinder rotation is of the order of $St=fD/U_{\infty}=1$, where $f$ is the dimensional frequency. The open-loop rotation of the cylinder needs to be this powerful in order to change the wake's stability properties: the train of small vortices that are generated in each shear layer by the rotation and advected by the free-stream enhances the wake symmetry and suppresses the interaction between the shear layers. The associated form drag is therefore reduced significantly, but unsurprisingly, this actuation method uses more energy than it saves\cite{Bergmann2006c,Protas2002}.

He {\it et al.}\cite{He2000a} were the first to consider the optimal control of a cylinder wake at low Reynolds numbers using rotary oscillation. In their study, the optimal control was evaluated starting from the best open-loop sinusoidal configuration ($\phi=3$, $St=0.74$ for $Re=200$, where $\phi$ is the amplitude of the sinusoidally oscillating tangential surface velocity). They chose to find the periodic control waveform (with a several frequency components) that would minimize drag, based on a horizon of 3 (high-frequency) forcing periods, and a Reynolds number of 200 and 1000. The resulting waveform is qualitatively similar to the open-loop one (high amplitude and high frequency), but with marginally improved drag reduction. Homescu {\it et al.}\cite{Homescu2002} used a similar setup to minimize the difference between the flow state and a target flow-field, which was chosen to be the flow around the cylinder at $Re=2$. They considered Reynolds numbers of 100 and 1000 and chose to find either the optimal constant rotation rate or sinusoidal rotation amplitude of the cylinder, over horizons of up to 5 time-units (i.e. roughly one vortex shedding period). The optimal sinusoidal forcing parameters ($\phi=3.25$, $St=1.13$, for $Re=100$) are different from the ones found by He {\it et al.}\cite{He2000a} but of the same order of magnitude. Protas {\it et al.}\cite{Protas2002a} also studied this problem but considered a free waveform for the control and minimized the sum of the control power and the power required to overcome drag, for Reynolds numbers of 75 and 150. They considered horizons of up to roughly one vortex shedding period and reported not obtaining significant improvements by further increasing the length of the horizons. The resulting control waveform is discontinuous between each horizon so no clear physical mechanism was extracted to explain how the rotation affects the flow-field. Nevertheless, the drag is reduced by 7\% at $Re=75$ and 15\% at $Re=150$, and with a positive energy balance. On the other hand, the amplitude of the control waveform does not seem to be changing significantly over time and the tangential cylinder velocity is typically of the order of 0.05 to 0.1 for $Re=75$ and of 0.1 to 0.2 for $Re=150$. Finally, Bergmann {\it et al.}\cite{Bergmann2005a,Bergmann2008} used reduced order models based on a Proper Orthogonal Decomposition of the flow-field to study the same problem. They chose to minimize the turbulent kinetic energy in the flow-field and the mean drag of the cylinder at a Reynolds number of 200. As the cylinder rotation was constrained to be sinusoidal, the control converged towards the energy inefficient open-loop optimum.

In the present investigation, a similar setup as the one in Protas {\it et al.}\cite{Protas2002a} is selected, but significantly longer control horizons than in all known earlier studies are considered (up to 100 convective time-units). In Sec.~\ref{numerics}, we introduce the numerical approach and problem setup used for the adjoint-based optimizations. In  Sec.~\ref{results}, we show the effectiveness of simply increasing the horizon length by choosing the same Reynolds number ($Re=75$) and cost function (based on total required power) as Protas {\it et al.}\cite{Protas2002a}, but with horizons of $50$ convective time-units. We then show that convincing results can also be obtained with different cost functions by minimizing squared lift or squared drag with an unconstrained control waveform, at a range of Reynolds numbers. The physical mechanism behind the performance of the optimal control is then analyzed. Finally, the influence of the cost function, of the length of the optimization horizon (10 to 100 convective time-units), and of the Reynolds number (between 75 and 200) are all discussed.

%%%

\section{Problem formulation and numerical method}\label{numerics}

The two-dimensional incompressible Navier-Stokes simulations presented in this article were run using a finite-volume immersed-boundary fractional step algorithm developed by Taira and Colonius\cite{Taira2007,colonius2008}. For this study, adjoint solvers were developed based on earlier work of Joe {\it et al.}\cite{Joea,JoeThesis} and Ahuja \& Rowley \cite{Ahuja2010}. In this section we provide an overview of the setup and procedures used and include further details about the solvers in Appendix~\ref{IBFS} and about the adjoint equations in Appendix~\ref{adjointequations}.

\begin{figure}
\includegraphics[trim=0cm 0cm 0cm 0cm, clip=true, width=0.9\textwidth]{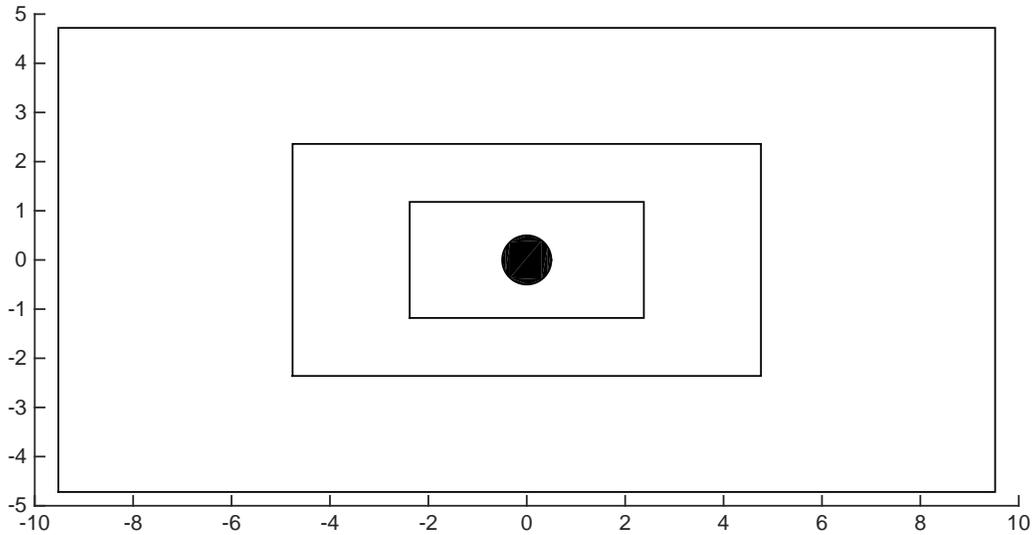}
\caption{Location of the cylinder within the first 3 nested grid levels (out of 8). \label{gridlevs}}
\end{figure}

\begin{figure}
\includegraphics[trim=0cm 0cm 0cm 0cm, clip=true, width=0.49\textwidth]{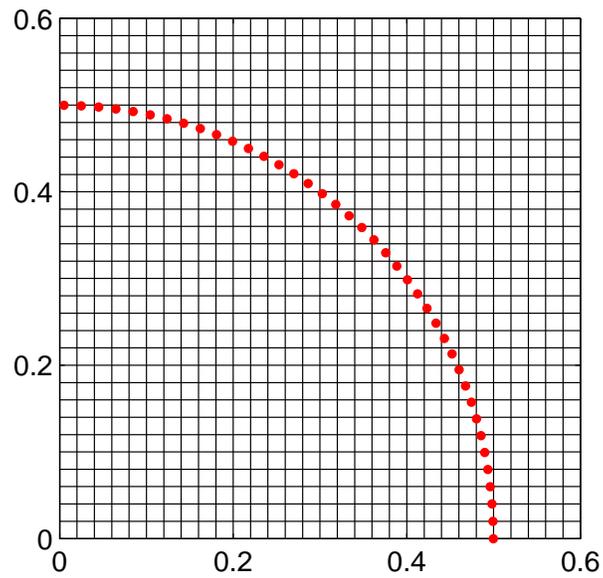}
\caption{(Color Online) Uniform Cartesian mesh of the finest grid level, in the vicinity of the immersed-boundary points (red dots) used to define the cylinder surface (of which a quarter is shown here). \label{cylindergrid}}
\end{figure}

A nominal set of parameters was used in all simulations presented in this article. The adequacy of these parameters was therefore checked as summarized in Table~\ref{gridconvtable}. Here, the vortex-shedding Strouhal number and mean drag coefficient of a stationary two-dimensional circular cylinder in a free-stream with a Reynolds number between 75 and 200 were computed with the nominal parameters, as well as with a more refined set of parameters, and the results were compared to those found in the literature. A good convergence and agreement with previous studies is obtained with the nominal set of parameters for both Reynolds numbers. For the nominal (refined) runs the time-step is $0.002$ ($0.001$) and the flow is solved on 8 nested grids of sequentially increasing size and correspondingly decreasing resolution (as introduced in Appendix~\ref{IBFS}). Each grid is composed of $240\times120$ ($600\times300$) cells and the extent of the smallest grid is $4.8\times2.4$ ($6.0\times3.0$) cylinder diameters. A plot of the cylinder's location within the first 3 (out of 8) nested grids levels in shown in Fig.~\ref{gridlevs}. As the adjoint solution is advected in the opposite direction to the forward one, the cylinder is located at the center of each grid level. The immersed-boundary points that define the cylinder is shown in Fig.~\ref{cylindergrid}, where it is superimposed on part of the mesh of the finest grid level.

\begin{table}
\caption{Strouhal number of vortex shedding and mean drag coefficient of the unforced flow around a stationary circular cylinder with the nominal and refined set of parameters, compared with values obtained in previous studies.}\label{gridconvtable}
\begin{ruledtabular}
\begin{tabular}{lccc}
  Simulation                               &$Re$ & Strouhal Number & Mean $C_D$ \\\hline
  Nominal                                  &100  &0.163            &1.330       \\
  Refined                                  &100  &0.164            &1.327       \\\hline
  Braza {\it et al.}\cite{Braza1986}        &100  &0.16             &1.36       \\
  Henderson\cite{Henderson1997}            &100  &0.166            &1.350       \\
  Park {\it et al.}\cite{Park1998}         &100  &0.164            &1.33       \\
  He {\it et al.}\cite{He2000a}            &100  &0.167            &1.353       \\  
  \hline\hline
  Nominal                                  &200  &0.194            &1.327       \\
  Refined                                  &200  &0.196            &1.356       \\\hline
  Braza {\it et al.}\cite{Braza1986}        &200  &0.20             &1.39       \\
  Henderson\cite{Henderson1997}            &200  &0.197            &1.341       \\
  He {\it et al.}\cite{He2000a}            &200  &0.198            &1.356       \\
  Bergmann {\it et al.}\cite{Bergmann2005a}&200  &0.200            &1.390         
\end{tabular}
\end{ruledtabular}
\end{table}

The adjoint-based optimization procedure considered in the present work aims to minimize a cost function of the form ${\cal J}(x,\phi)=\int_0^T{\cal I}(x,\phi)dt$, which depends on the state of the flow $x(t)$ and control waveform $\phi(t)$ as defined in Appendix~\ref{adjointequations}. In this case, the integral of the power, the squared lift or the squared drag over the control horizon were used:
\begin{eqnarray}
{\cal J_P}&=&\int_0^T \left\{\left(Power\right)_{Drag}+\left(Power\right)_{Control}\right\}dt\label{eq:costP},\\
{\cal J_L}&=&\int_0^T \frac{1}{2}\left(Lift\right)^2dt\label{eq:costL},\\
{\cal J_D}&=&\int_0^T \frac{1}{2}\left(Drag\right)^2dt\label{eq:costD},
\end{eqnarray}
The cost function is minimized using a single degree of freedom but time-dependent control $\phi(t)$: The tangential velocity of the cylinder surface. Note that there is no control penalization term in (\ref{eq:costL}) and (\ref{eq:costD}). Although in general there is no guarantee that the problem will be well-posed without regularizing the control effort, it was found that for the setup presented here, no such term was necessary for long enough horizons (50 convective time-units or more), in order to obtain a converged optimal solution with a bounded control effort.

In order to locally minimize these cost functions, the gradient of the cost with respect to the controls is determined in the standard manner\cite{Bewley2001a, Joea}, as detailed in Appendix \ref{adjointequations}. Using calculus of variations, this procedure yields a set of conditions that must be satisfied in order to identify the optimal control. First, the spatially discretized version of the vorticity form of the Navier-Stokes equations (\ref{eq:nseq}) must hold throughout the simulation:
\begin{eqnarray}
\left\{
\begin{array}{l}
\dot \omega=\nabla\times\left( u\times\omega\right) -Re^{-1}\nabla\times\left(\nabla\times\omega\right)+\nabla\times f,\\
u_B=\phi(t) \hat u_t.
\end{array}\right.
\label{eq:nseq}
\end{eqnarray}
Here $u$ refers to velocity, $\omega=\nabla\times u$ to vorticity and $\dot{\omega}=\partial \omega/\partial t$, $u_B$ is the velocity on the body surface, and $\phi(t)$ is the control, which as mentioned above imposes a \textit{tangential} velocity along the body surface, mimicking cylinder rotation: $\hat u_t$ is a locally defined unit vector, which points in the direction that is tangential to the body surface in the counterclockwise direction. The immersed body force $f$ is used to impose the jet velocity along the body surface.

Second, the corresponding adjoint Navier-Stokes equations must also hold throughout the simulation. These can be shown to be a spatially discretized version of the following equations (further details in Appendix \ref{adjointequations}):
\begin{eqnarray}
 \left\{
 \begin{array}{l}
 -\dot \omega^+=\nabla\times\left( \omega \times u^+\right) -\nabla^2\left( u^+\times u\right)-Re^{-1}\nabla\times\left(\nabla\times\omega^+\right)+\nabla\times f^+,\\
 u_B^+=u^+_{slip},
 \end{array}
 \right.
 \label{eq:adjeq}
 \end{eqnarray}
where $^+$ refers to adjoint quantities. $u^+_{slip}$ is the adjoint ``slip velocity'' at the body surface. The adjoint equations are forced through this term, and this forcing depends on the cost function (as again shown in Appendix~\ref{adjointequations}).

Finally, the gradient of the augmented cost function with respect to the controls ${\cal G}$ must be 0. The gradient can readily be computed given the data from the forward and adjoint simulations. However, $\phi(t)$ is in general not optimal so ${\cal G}$ is in general \textit{non-zero}. It is instead used to iteratively update the control waveform using a conjugate gradient approach.

The full adjoint optimization procedure can thus be outlined as follows:
\begin{enumerate}
\item A cost function ${\cal J}(x,\phi)$, an initial control guess $\phi(t)$, and a starting condition $x(0)$ are defined;
\item The Navier-Stokes equations are solved in order to evaluate $x(t)$ from $t=0$ to $t=T$, starting from the initial condition $x(0)$, and with the current control guess $\phi(t)$. The corresponding cost ${\cal J}(x,\phi)$ is also evaluated;
\item The adjoint equations are solved backwards in time, in order to evaluate $x^+(t)$ from $t=T$ to $t=0$, with the ``initial'' condition $x^+(T)=0$. The control gradient ${\cal G}$ is calculated from the results of the forward and adjoint simulations;
\item The optimal control update \textit{distance} is calculated iteratively based on the calculated gradient ${\cal G}(x,x^+,\phi)$, using a line minimization algorithm (more details below);
\item The control guess $\phi(t)$ is updated according to the results of the line minimization, allowing the new cost to be evaluated;
\item Steps 2. to 5. are repeated using the updated control guess $\phi(t)$ until convergence;
\item The starting time is advanced by about $2/3$ of the control horizon (in this case) and the state of the controlled flow-field at that time is used as the new initial condition. The last third of the converged optimal control signal is discarded. The new initial control guess is set to zero (more details below);
\item Steps 1. to 7. are repeated for the next horizon. 
\end{enumerate}
More details about this procedure are included for instance in \cite{Bewley2001a,Joea,Protas2002a,JoeThesis}.

The line minimization algorithm mentioned in the 4th step above is necessary as the control gradient only provides information about the \textit{direction} in which the control needs to be updated in order to optimally reduce the cost function (locally). A further inner iteration is then required to find the optimal {\it magnitude} of the control update in the gradient direction. A steepest gradient algorithm can be used, but it is more efficient to use a conjugate gradient approach\cite{Press2007}. Typically, a Brent line minimization is then used to find the optimal update distance\cite{Press2007}. However, in order to make this searching procedure more efficient, a generalized version of Brent's search algorithm was developed and used in the present work, whereby several forward simulations are run together in parallel.

In order to obtain optimization results that are longer than one horizon, a new initial control guess and starting condition are necessary to start the optimization at each horizon. As mentioned in Step 7. above, the initial condition for each horizon $i$ of total length $T_i$ was chosen to be a snapshot of the converged solution of the previous horizon $i-1$ such that $t_{i}(0)\approx(2/3)T_{i-1}$. A large enough fraction of the optimal control solution should be discarded to ensure transients from the adjoint simulation do not affect the retained solution. Here, keeping about $2/3$ of the optimal control was chosen as an adequate value to truncate the part of the solution that was most affected by adjoint transients. Reasonably small variations in this value would only be expected to change the performance of the control slightly. As it was chosen to set the first control guess to zero rotation throughout each new horizon, an initial discontinuity is to be expected in the optimal control signal since ${\phi((2/3)T_{i-1})\neq0}$ in general.

The gradients computed with this code are not the exact discrete gradients of the forward problem. Indeed, the time-stepping scheme is not exactly self-adjoint, as the base-flow is unsteady and the same multi-step marching method is used for the forward and adjoint solvers, as described in Appendix~\ref{IBFS}. Additionally, the same nested grid procedure was used in the forward and adjoint solvers, but this procedure is in fact not self-adjoint either, as noted by Ahuja and Rowley\cite{Ahuja2010}. Finally, solving the adjoint equations requires storing the entire solution of the forward problem (the unsteady base-flow) since $u$ and $\omega$ appear in the advection terms of Eq.~(\ref{eq:adjeq}). To reduce the computational requirements of this procedure, we only store every 10th snapshot of the forward solution and use linear interpolation to reconstruct it at a cheaper cost. The error associated with this procedure is very small, as the base-flow evolves at a much slower rate than the time-step. For instance, at $Re=100$, using interpolation introduces a relative error in the gradient that is typically of the order of $0.001\%$ to $0.005\%$ (with a control guess of $\phi(t)=0$ and a horizon of $10$ convective time-units). Nevertheless, as a consequence of the points mentioned above, we can only expect to obtain an approximation to the exact discrete optimal solution of the problem.

In order to check that the gradients obtained with the code are of an acceptable accuracy, a standard finite-difference check\cite{Homescu2002} was performed on the adjoint code: if the forward and adjoint equations are satisfied, then the following Taylor expansion of the cost function can be written:
\begin{eqnarray}
{\cal J}(\phi)- {\cal J}(\phi+\delta\phi)=\int_0^T{\left({\cal G}^T\delta\phi +O(\delta\phi^2)\right)dt},\label{eq:gradcheck}
\end{eqnarray}
where ${\cal G}(t)$ is the control gradient. The nominal control is chosen to be $\phi(t)=0$ and $\delta\phi(t)$ is an arbitrary control perturbation, chosen to be a sinusoidal function of the form ${\delta\phi(t)=\epsilon \sin{\left(\left(2\pi St\right) t\right)}}$, for small enough $\epsilon$. Here $\epsilon=10^{-6}$ was used and Strouhal numbers in the range $0.1\le St\le 0.5$ were tested. The error between the two sides of Eq.~(\ref{eq:gradcheck}) normalized by $\left|{{\cal J}(\phi)-{\cal J}(\phi+\delta\phi)}\right|$ was checked and found to be typically of the order of $0.5\%$, $1\%$ and $3\%$ or less for $T=10$, $T=50$, and $T=100$, respectively for all three cost functions given in Eq.~(\ref{eq:costP}, \ref{eq:costL}, \ref{eq:costD}) and at $Re=100$. Due to the fact that the error seems to increase with $T$, we can expect to obtain more suboptimal results with longer horizons.

%%%

\section{Results and discussion}\label{results}

\begin{table}
\caption{Summary of results from the optimizations: the horizon length is given in convective time-units, ${\cal J}$ refers to the cost functions defined in Eq.~(\ref{eq:costP}, \ref{eq:costL}, \ref{eq:costD}). $Re$ is the Reynolds number, $C_D$ and $C_L$ are the drag and lift coefficients, and $\phi$ is the control amplitude. The values quoted here refer to the reduction in mean $C_D$ and the relative change in $C_L$ amplitude (with respect to the unforced case at the same $Re$), as well as the amplitude of the tangential velocity $\phi$. All values are estimated towards the end of the last considered horizon for each Run, ignoring the part of the response where adjoint transients become significant. The results are also compared to those obtained by Protas {\it et al.}\cite{Protas2002a}.\label{resultstable}}
\begin{ruledtabular}
\begin{tabular}{ccccccc}
  Run&Horizon&${\cal J}$&$Re$&$C_D$&$C_L$&$\phi$ \\\hline
  Protas {\it et al.}\cite{Protas2002a}&6&${\cal J_P}$&$75$&$7\%$&--&$0.05-0.1$\\
  Protas {\it et al.}\cite{Protas2002a}&6&${\cal J_P}$&$150$&$15\%$&--&$0.1-0.2$\\\hline
  1&2$\times$50&${\cal J_P}$&$75$& $13\%$&$-99\%$&$0.0007$\\
  2&2$\times$50&${\cal J_D}$&$75$& $13\%$&$-95\%$&$0.002$\\\hline
  3&6$\times$10&${\cal J_D}$&$100$&$>60\%$&$>+1900\%$&$>2.0$\\
  4&50&${\cal J_D}$&$100$&$16\%$&$-89\%$&$0.04$\\
  5&2$\times$50&${\cal J_D}$&$100$&$19\%$&$-98\%$&$0.001$\\
  6&100&${\cal J_D}$&$100$&$10\%$&$-69\%$&$0.10$\\
  7&50&${\cal J_L}$&$100$&$6\%$&$-73\%$&$0.07$\\\hline
  8&50&${\cal J_D}$&$200$&$9\%$&$-58\%$&$0.21$\\
  9&2$\times$50&${\cal J_D}$&$200$&$9\%$&$-46\%$&$0.14$\\
  10&100&${\cal J_D}$&$200$&$7\%$&$-38\%$&$0.12$\\
  11&50&${\cal J_L}$&$200$&$11\%$&$-64\%$&$0.21$
\end{tabular}
\end{ruledtabular}
\end{table}

Several optimizations were run in order to investigate the influence of changing the cost function, the horizon length, and the Reynolds number, as summarized in Table \ref{resultstable}. The performance of the optimizations is quantified by the amount drag reduction and suppression of lift fluctuations obtained towards the end of the optimized simulations.

As shown in Runs 1 and 2, in which the Reynolds number was chosen to be the same as the one used by Protas {\it et al.}\cite{Protas2002a} ($Re=75$), improved results are obtained just by increasing the length of the control horizon, both when the same cost function is used (Run 1 with ${\cal J_P}$) and when the drag cost function is used (Run 2 with ${\cal J_D}$). Table~\ref{resultstable} shows that only a small amount of control is required towards the end of the horizon, and the lift fluctuations and mean drag are both reduced by a larger amount than with the short horizons used in previous work.

In Sec.~\ref{optimalcontrolanalysis}, we investigate how the optimal control leads to such large reductions in the mean drag and lift fluctuations, by focusing on Run 5, where the Reynolds number is increased to $Re=100$ and ${\cal J}={\cal J_D}$. We then discuss the influence of the cost function, the Reynolds number and the horizon length on the overall optimization performance in Sec.~.\ref{parametercomparison}.

\subsection{Suppression of vortex shedding}\label{optimalcontrolanalysis}

\begin{figure}
\includegraphics[trim=0.2cm 0.8cm 0cm 0cm, clip=true, width=0.49\textwidth]{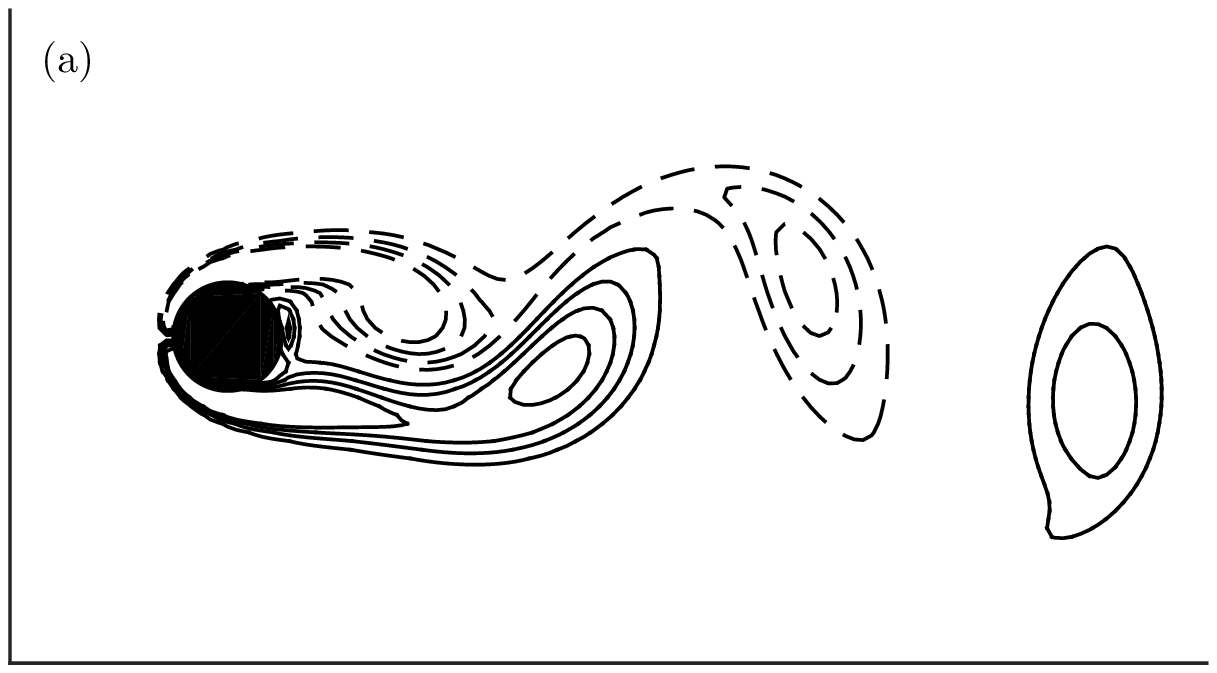}
\includegraphics[trim=0.2cm 0.8cm 0cm 0cm, clip=true, width=0.49\textwidth]{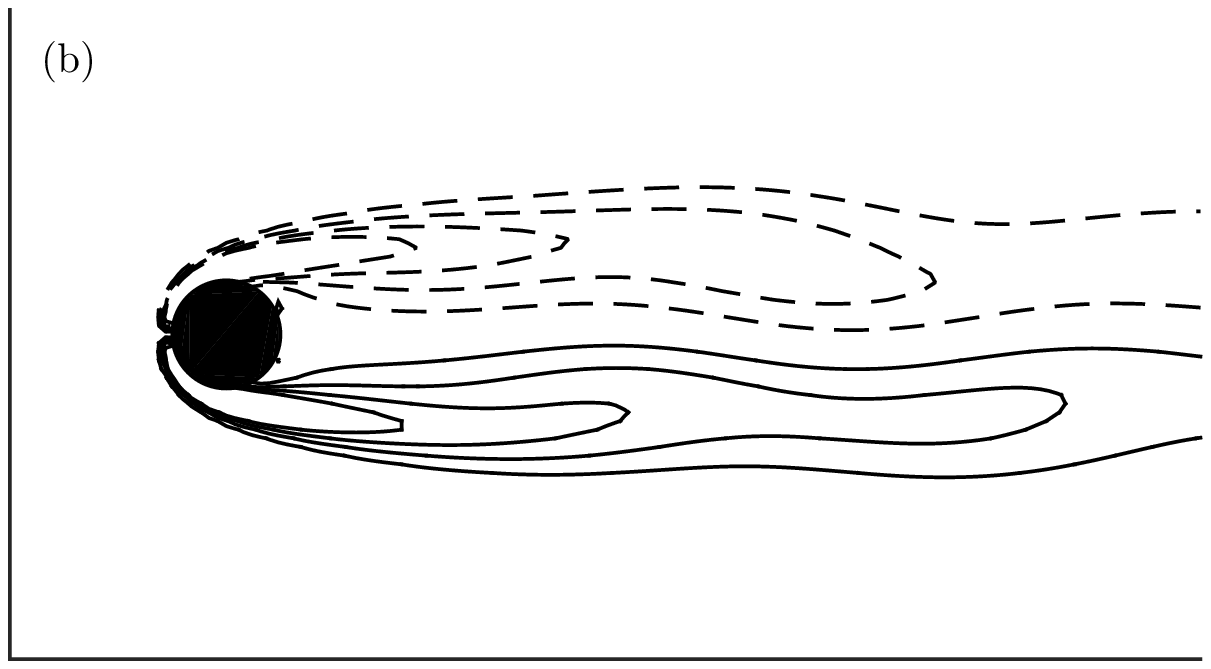}
\caption{Results from Run 5. (a) Unforced wake vorticity contours. (b) Optimally controlled wake vorticity contours. Solid (dashed) lines represent positive (negative) vorticity. Contours shown for vorticity values of -2 to -0.5 and 0.5 to 2 in increments of 0.5. The vortex shedding is almost fully suppressed in the controlled case.\label{optcontrolflow}}
\end{figure}

\begin{figure}
\includegraphics[trim=0cm 0cm 0cm 0cm, clip=true, width=0.9\textwidth]{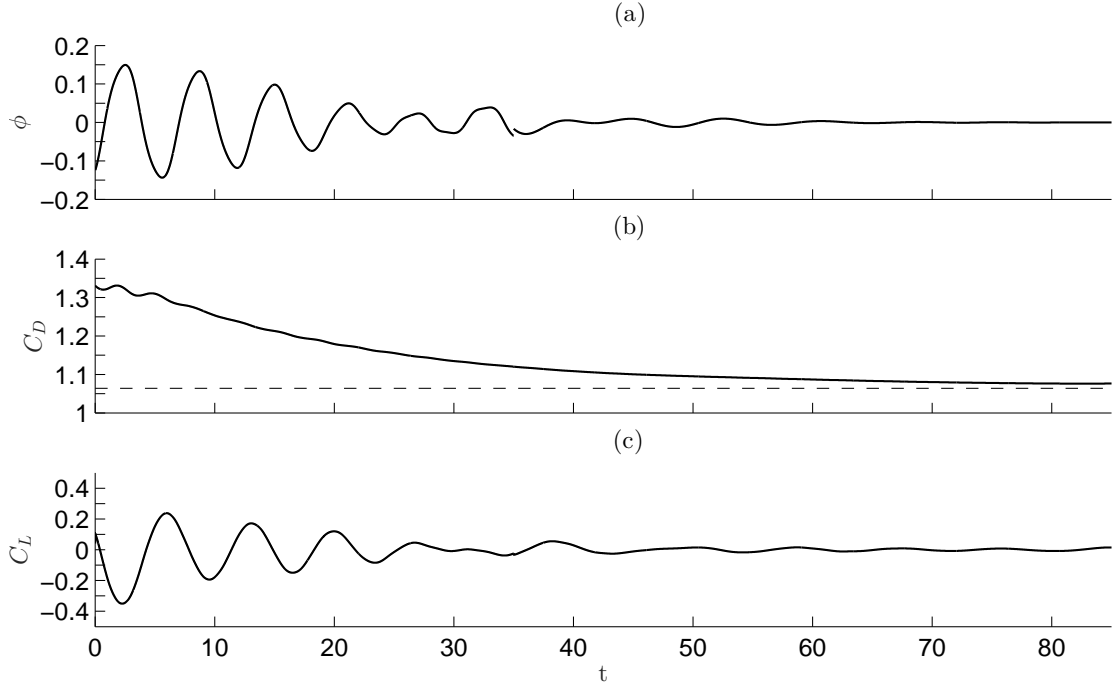}
\caption{Results from Run 5. (a) Optimal control waveform and resulting drag (b) and lift (c) coefficients, showing a clear suppression of the lift oscillations, a significant reduction in the mean drag, and a control waveform that tends to nearly zero amplitude for large times. In (b) the drag of the unstable equilibrium of the flow-field, computed by using Selective Frequency Damping\cite{Jordi2014,Akervik2006}, is also plotted as a reference (dashed line).\label{optcontrolforces}}
\end{figure}

In Fig.~\ref{optcontrolflow} the unforced flow-field (a) is compared to the optimally controlled flow-field (b) of Run 5. The intensity of the vortex shedding in Fig.~\ref{optcontrolflow}(b) is reduced and the flow appears to be much more symmetric. The optimal control signal is shown in Fig.~\ref{optcontrolforces}(a), while the optimally controlled drag and lift coefficient signals are shown in Fig.~\ref{optcontrolforces}(b) and (c) respectively. In Fig.~\ref{optcontrolforces}(a), the drag of the unstable (steady) equilibrium is also plotted (this value was computed by using Selective Frequency Damping\cite{Jordi2014,Akervik2006}). Clearly, the controlled flow-field has been stabilized to a drag state that is close to that of this unstable equilibrium.

From Fig.~\ref{optcontrolflow}, Fig.~\ref{optcontrolforces}, and the values from Table \ref{resultstable}, it is clear that in Run 5 the vortex shedding is almost fully suppressed and that the lift oscillations and the mean drag are both reduced significantly. Moreover, towards the end of the second optimization window, the tangential velocity on the cylinder surface is of the order of $0.1 \%$ of the incoming flow velocity. Comparing these results to previous work with similar setups, only Protas {\it et al.}\cite{Protas2002a} obtained comparable drag reductions without needing an excessively large amount cylinder rotation. However, the control waveform obtained by Protas {\it et al.}\cite{Protas2002a} is discontinuous due to the chosen setup for the optimizations: only short optimization horizons were used. In the present case, almost no cylinder rotation is required to keep the Von K\'{a}rm\'{a}n wake suppressed for large times. Moreover, we obtain a smooth control waveform, which clearly has a main frequency component that corresponds to the vortex shedding frequency. We therefore proceed to investigate in what way the wake is affected by the control.

\begin{figure}
\includegraphics[trim=0cm 0cm 0cm 0cm, clip=true, width=0.9\textwidth]{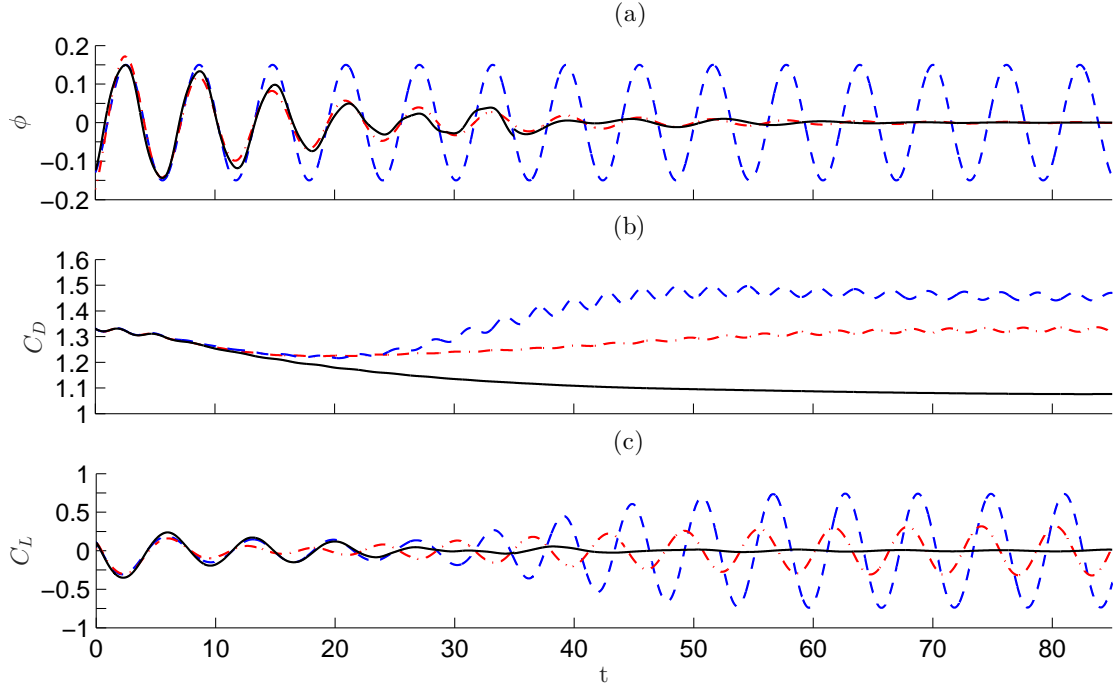}
\caption{(Color Online) Results using harmonic approximations to the optimal control waveform of Run 5 (also run with $Re=100$): (a) Control signals and resulting (b) drag and (c) lift coefficients. Black solid line: optimal control. Red dash-dotted line: exponentially damped sinusoidal forcing. Blue dashed line: sinusoidal forcing. Despite a similar initial behavior, neither of the open-loop approximations are able to suppress the shedding sustainably.\label{harmonicappxsig}}
\end{figure}

In previous optimal control studies\cite{He2000a,Homescu2002,Bergmann2008}, the aggressive optimal control obtained can be seen as essentially an open-loop strategy with optimized  (constant) parameters: purely harmonic forcing cannot suppress the wake instability, it can only alter it to a more favorable drag state with suitably large rotational velocities. In order to test whether the low drag state reached with optimal control can also be obtained with similar open-loop control signals, two open-loop control waveforms that approximate the optimal control were designed, as shown in Fig.~\ref{harmonicappxsig}: the first is a sinusoidal signal whose frequency, amplitude, and phase approximately match the optimal control at $t=0$ and the second is an exponentially decaying sinusoidal waveform (also with approximately matching frequency, amplitude and phase at $t=0$). Upon applying these control signals to the cylinder, the drag is initially reduced in both cases, but it returns to its unforced value for the exponentially decaying sinusoidal case. In the sinusoidal case the control actually \textit{increases} the drag by $9 \%$ for large times, with respect to the unforced case. This type of behavior is common: the vortex shedding frequency naturally locks-in to the forcing frequency when it is close to the unforced shedding frequency, thereby amplifying the vortex shedding intensity (e.g. \cite{Griffin1991}).

These simple tests therefore seem to confirm that there is an intrinsic need for feedback information to be available to the control for it to damp the vortex shedding effectively. Thiria\cite{Thiria2006} studied the influence of the rotation of the cylinder on the wake vorticity in the {\it open-loop} case and argued that when the optimal drag reduction forcing parameters are chosen, the rotation of the cylinder has a ``destructive'' influence on the formation of vortices, since the cylinder rotation creates vorticity of the {\it opposite} sign to that of the vortex being formed. At the same time, the rotation promotes the creation of a vortex in the opposite shear layer, thus effectively enhancing the synchronization of the shedding of vortices in the wake. On the other hand, when the rotation is locked-in with the vortex shedding, the interaction can be seen as ``constructive'' since the cylinder rotation creates vorticity of the {\it same} sign as that of the vortex being formed and reduces the amount of vorticity in the opposite shear layer, thus aggravating the shear layer interaction and strength of the vortices.

\begin{figure}
\includegraphics[trim=0.2cm 0.75cm 0cm 0cm, clip=true, width=0.49\textwidth]{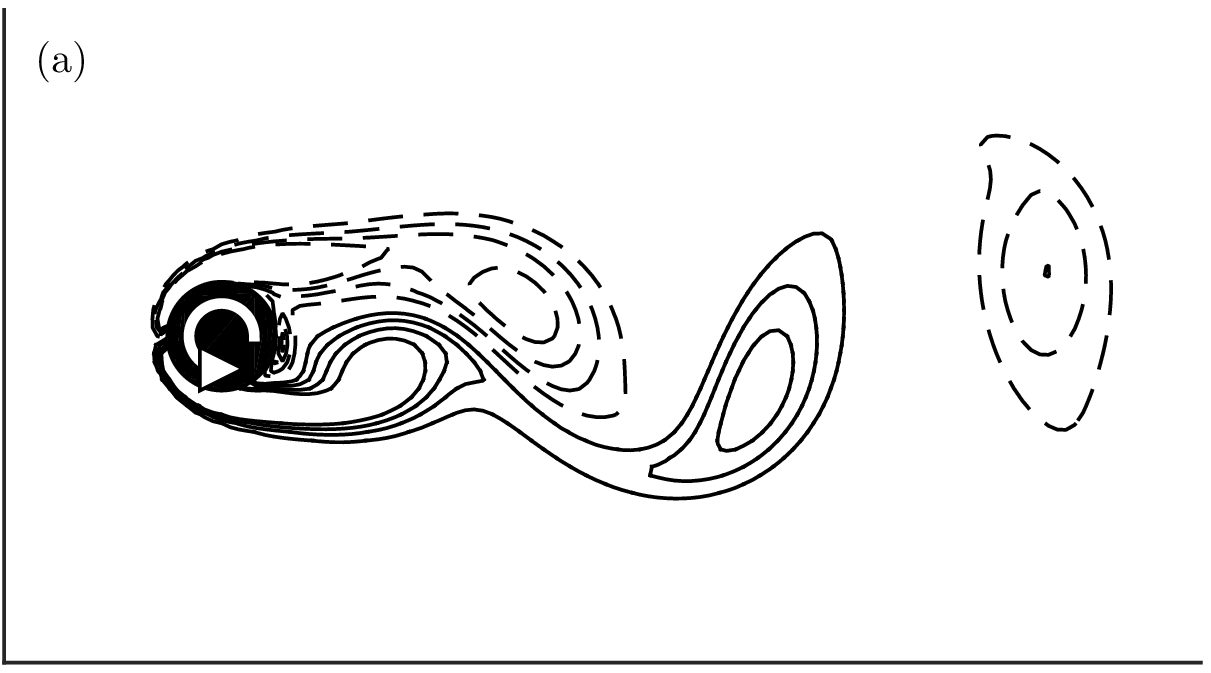}
\includegraphics[trim=0.2cm 0.75cm 0cm 0cm, clip=true, width=0.49\textwidth]{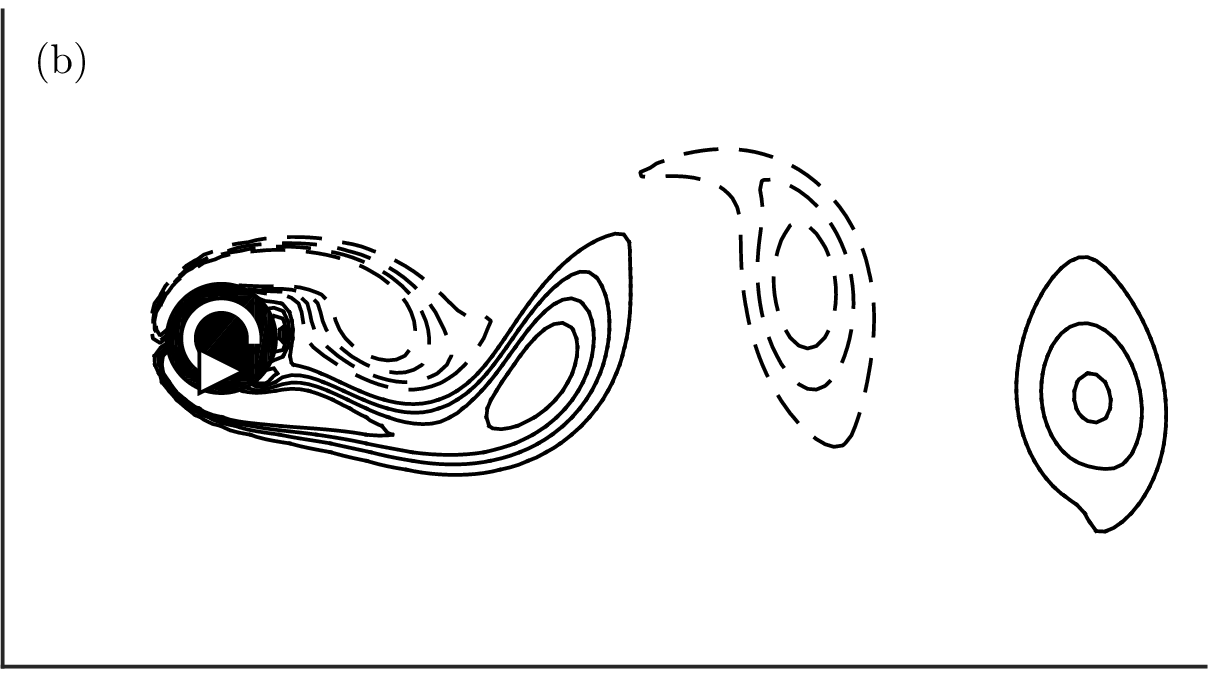}\\
\includegraphics[trim=0.2cm 0.75cm 0cm 0cm, clip=true, width=0.49\textwidth]{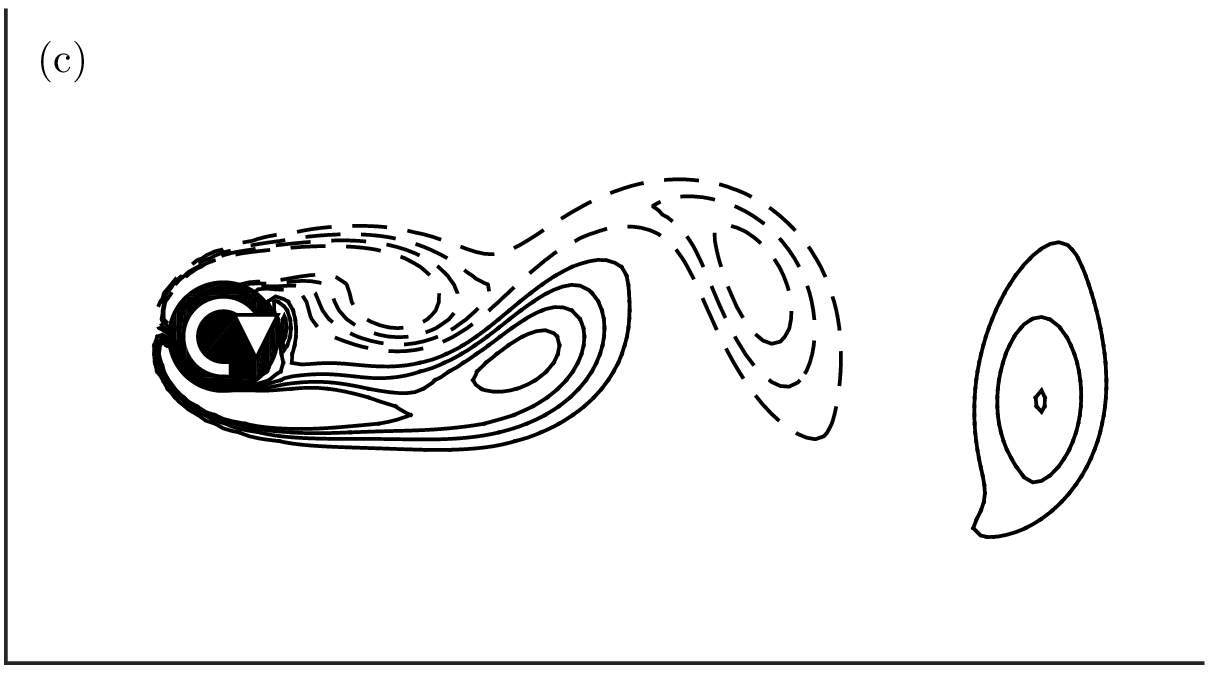}
\includegraphics[trim=0.2cm 0.75cm 0cm 0cm, clip=true, width=0.49\textwidth]{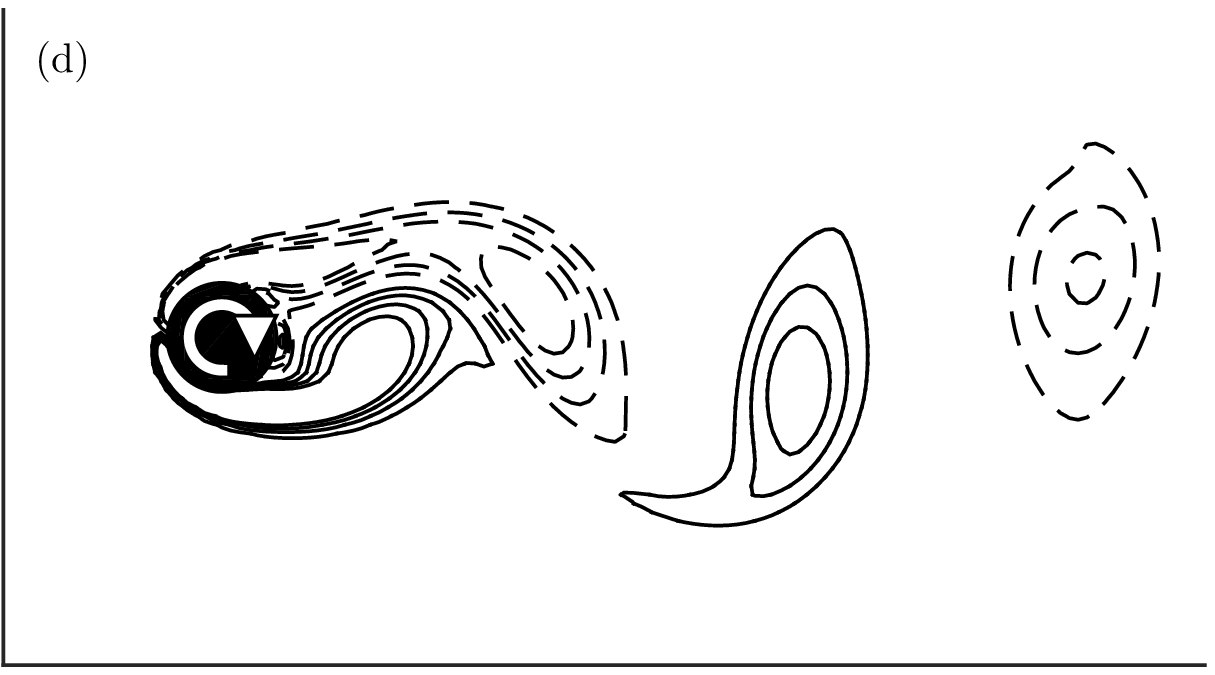}
\caption{Vorticity contours for the sinusoidal forcing case with amplitude, frequency and phase matching the optimal control for early times. Solid (dashed) lines represent positive (negative) vorticity. Contours shown for vorticity values of -2 to -0.5 and 0.5 to 2 in increments of 0.5. (a,b) Maximal value of positive (anti-clockwise) rotation. (c,d) Maximal value of negative (clockwise) rotation. (a,c) In the first few shedding periods, the cylinder rotation is ``destructive'' and reduces the intensity of the shedding. (b,d) Once lock-in occurs, the rotation becomes ``constructive'' and reinforces the shedding.\label{harmonicappxflow}}
\end{figure}

Comparing the direction of the rotation with the vortex shedding phase for the open-loop sinusoidal case described above (blue dashed line in Fig.~\ref{harmonicappxsig}) the same conclusions can be drawn: as shown in Fig.~\ref{harmonicappxflow}(a) and (c), for $t<10$, the rotation is in the ``destructive'' direction, while for $t>60$, (Fig.~\ref{harmonicappxflow}(b) and (d)) the vortex shedding is locked-in and the rotation is in the ``constructive'' direction. This suggests that matching the cylinder's rotation direction to the phase of the vortex shedding is a promising strategy to suppress the wake fluctuations without requiring the full optimal control framework.

One way to interpret this conclusion is that the mechanism leading to the suppression of vortex shedding in the open-loop experiments of Tokumaru and Dimotakis\cite{Tokumaru2006a} and the subsequent related studies is effectively the same as the one allowing suppression of vortex shedding in the energy-efficient optimal control case shown here. In the open-loop case, a large amount of input energy is required in order for the interaction between the destructive sinusoidal rotation and the vortex shedding to become stable. On the other hand, similar performance is possible with a much smaller control effort if information is fed back to the controller as this allows the unstable destructive interaction to be maintained without requiring the fast high-amplitude rotation that is necessary with open-loop forcing.

\subsection{Impact of optimization and simulation parameters}\label{parametercomparison}

The results in Table \ref{resultstable} show that at a Reynolds number of 100, minimizing the RMS drag is more efficient in this case than minimizing the RMS lift (see Run 4 compared to Run 7). For a Reynolds number of 200 however, the performance is comparable. For both cost functions and Reynolds numbers, the cylinder rotation is slow, of small amplitude and roughly at the vortex shedding frequency, and the drag is significantly reduced due to the suppressed intensity of vortices in the wake. In previous adjoint optimization studies (e.g. \cite{Min1999,Bewley2001a}) it was also found that it is not always immediately obvious why one cost function performs better than another one, due to the highly non-convex and nonlinear nature of the problem. Clearly here, we expect both ${\cal J_L}$ and ${\cal J_D}$ to be minimized if the vortex shedding is somehow completely suppressed, but there is no guarantee that using the two cost functions will in fact lead to similar modifications in the flow-field, as the path to the local minimum reached and even the minimum itself may be different. The present results therefore bring a further confirmation that it is worth running optimizations with several cost functions even if all of them are expected to be reduced  by the desired flow behavior. Note that although all simulations were run for enough iterations to appear converged, it is possible that further reductions in the cost could be obtained by computing a larger number of iterations, since changing ${\cal J}$ also affects the convergence properties of the optimization problem.

Table \ref{resultstable} also shows that increasing the Reynolds number from $Re=100$ to $Re=200$ is overall detrimental to the relative performance of the control. From linear stability analysis\cite{Huerre1990}, the real part of the unstable eigenvalues associated with the vortex shedding grows with the Reynolds number and several studies\cite{Lauga2003a,Lu2011} also reported a reduction of closed-loop performance as the Reynolds number is increased. In the present case, the two shear layers are more unstable for $Re=200$ than $Re=100$, so even if  the cylinder rotation is able to initially enhance the symmetry of the wake and delay the interaction between the shear layers, the instability is not as readily suppressed for the entire wake region and the drag reduction is therefore not as significant. At both $Re=75$ and $Re=100$, the flow can be considered to be effectively stabilized (as suggested by Table~\ref{resultstable}, which shows that the lift fluctuations are almost fully suppressed). At $Re=75$ however, as the Reynolds number is lower than at $Re=100$, not only is the vortex shedding is less intense, but the unstable equilibrium also has a higher $C_D$ \cite{Bergmann2005a}. As a result, a smaller relative drag reduction is possible by stabilizing the wake in this case.

A key difference between previous work and the setup chosen here is that in this study, it was chosen to use much longer optimization horizons. Protas {\it et al.}\cite{Protas2002a} found that the horizon length should be longer than one vortex shedding period but only obtained marginal improvements by further increasing it. Bewley {\it et al.}\cite{Bewley2001a} argued that if the horizon is too long the optimization can become excessively non-convex and computationally expensive, leading to suboptimal performance in practice. Furthermore, inaccuracies in the computed gradient have the potential to accumulate and stall the optimization for very long horizons, leading to more suboptimal results. In this article, optimizations with horizon lengths of $10$, $50$ and $100$ convective time-units were run.

Considering the $100$ convective time-unit optimizations, a similar behavior to the $50$ convective time-unit case can be seen in Table \ref{resultstable}, but with an inferior performance (see Run 5 compared to Run 6 and Run 9 compared Run 10). This correlates well with the fact that 100 convective time-units is indeed an excessively long horizon in this case. As mentioned above, it is possible that additional iterations may still result in a further reduction in the cost. Nevertheless, the present results confirm that such long horizons can indeed result in prohibitively slow convergence of the optimization problem or suboptimal converged results.

\begin{figure}
\includegraphics[trim=0cm 0cm 0cm 0cm, clip=true, width=0.9\textwidth]{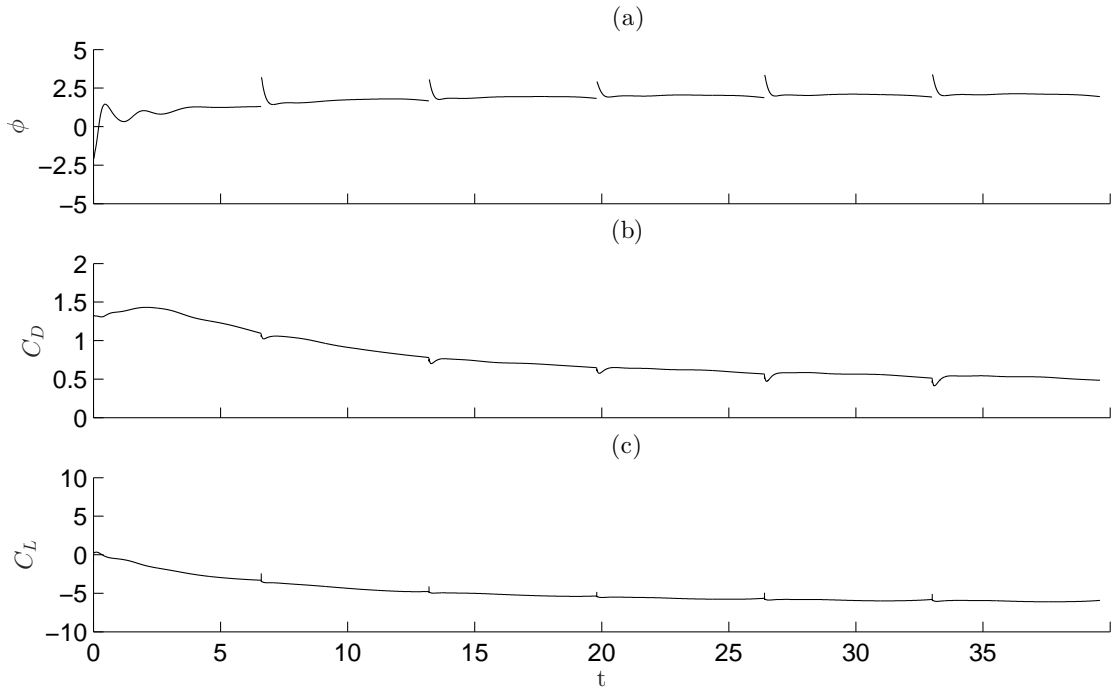}
\caption{Results from Run 3. (a) Optimal control waveform and resulting drag (b) and lift (c) coefficients. The optimal control leads to a non-zero-mean, fast rotation optimum.\label{horizon10}}
\end{figure}

\begin{figure}
\includegraphics[trim=0.2cm 0.75cm 0cm 0cm, clip=true, width=0.49\textwidth]{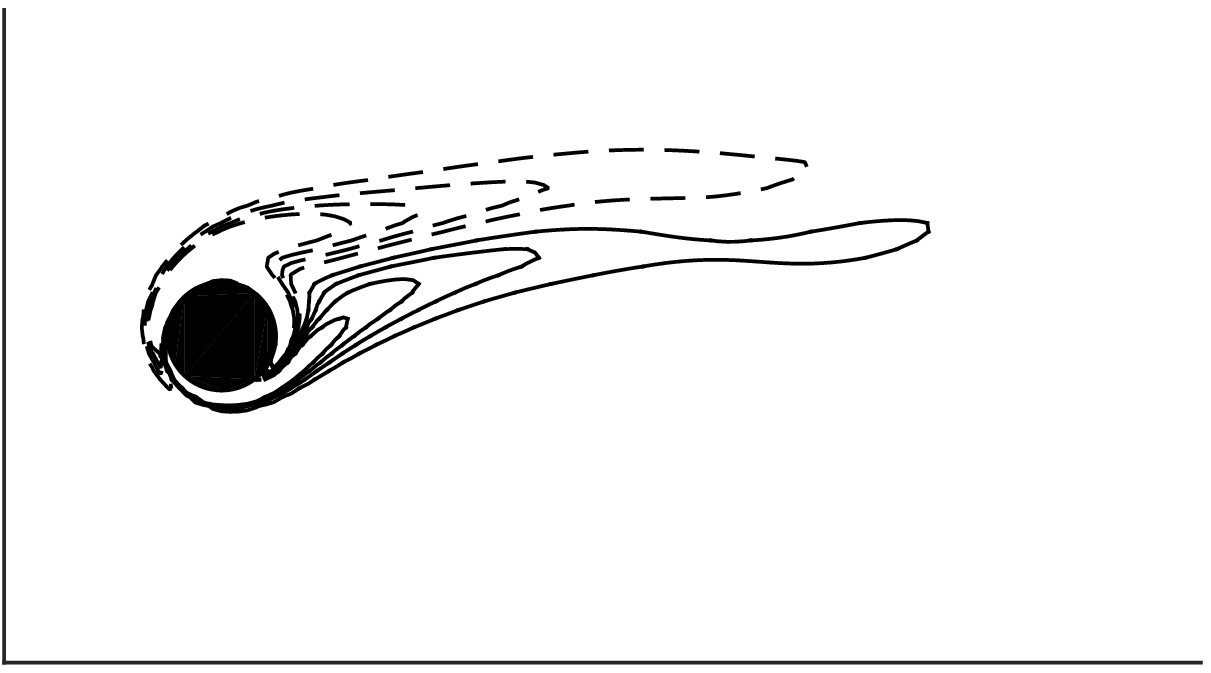}
\caption{Vorticity contours corresponding to Run 3. Solid (dashed) lines represent positive (negative) vorticity. Contours shown for vorticity values of -2 to -0.5 and 0.5 to 2 in increments of 0.5. The flow-field is strongly asymmetric due to the large nearly constant rotation of the cylinder, which leads to reduced drag.}\label{constant rot}
\end{figure}

When on the other hand the horizon length is $1$ convective time-unit (not shown here), i.e. shorter than a vortex shedding period, it was found that the problem does not converge at all, as it is ill-posed from the first optimization horizon, without a control penalty term. With a horizon length of $10$ convective time-units however (Run 3), the solution of the optimization tends towards a well-known low-drag flow-state with nearly {\it constant} rotation, as shown in Fig.~\ref{horizon10}. The control waveform has discontinuities at the start of each horizon but the general behaviour of the control is still clearly visible. Homescu {\it et al.}\cite{Homescu2002} and Kang {\it et al.}\cite{Kang1999} also showed that the cylinder drag can be reduced significantly and the vortex shedding suppressed completely by using sufficiently fast constant rotation of the cylinder in one direction. Unsurprisingly, this results in a significant constant lift force on one side of the cylinder, as shown in Fig.~\ref{horizon10}(c). Both studies found an optimal rotation rate at $Re=100$ of $\phi\approx1.85$. Here the rotation rate quickly increases until it reaches a value that is close to this ``optimum'', but then continues to slowly increase. After $40$ convective time-units, the rotation rate is $\phi\approx2.1$ and the cylinder is still slowly accelerating. It is expected that computing further horizons would lead to a continued increase in this quasi-steady rotation rate, since the drag has been shown to decrease further for even higher rotation rates\cite{Kang1999}. The physical mechanism reducing drag and stabilizing the wake in Run 3 is therefore  effectively an open-loop strategy, which does not require closed-loop control but does require a significant amount of input energy.

These results show that the choice of horizon length can have a crucial impact on the solution of the optimization and suggest the existence of another important timescale than the vortex shedding frequency. It seems reasonable to expect a control that efficiently suppresses vortex shedding to do so over a length of time that is longer than one vortex shedding period. For instance a control waveform that eventually fully stabilizes the wake will not necessarily lead to the maximum drag reduction in the first few convective time-units.

\begin{figure}
\includegraphics[trim=0cm 0cm 0cm 0cm, clip=true, width=0.9\textwidth]{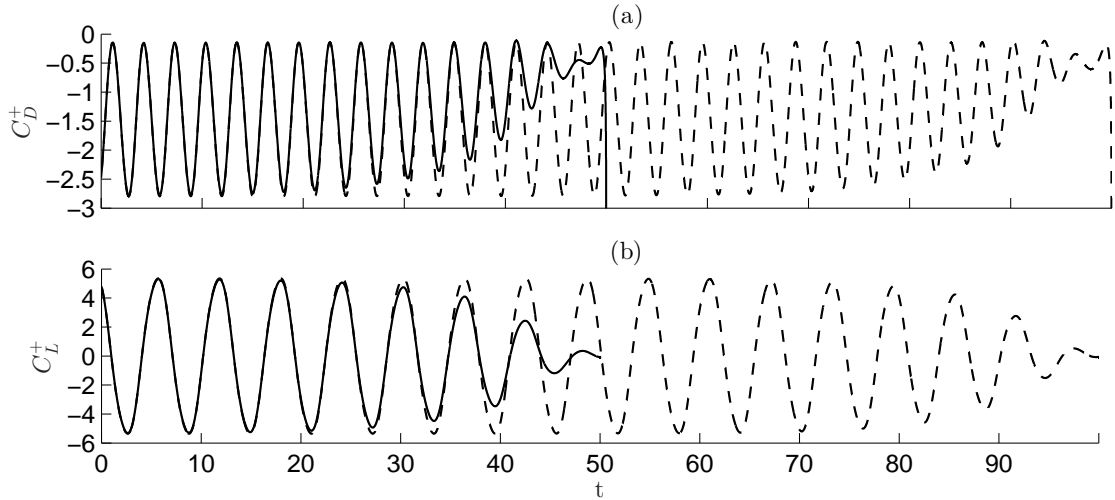}
\caption{Time history of the adjoint force coefficients for the first optimization iteration (no cylinder rotation in the corresponding forward run, i.e. $\phi(t)=0$) of Run 4 (solid line) and Run 6 (dashed line). (a) Adjoint drag $C_D^+$, (b) Adjoint lift $C_L^+$. Note that the adjoint force coefficients are defined in an analogous manner to the standard force coefficients (see Appendix~\ref{IBFS}), but based on the \textit{adjoint} forces at the immersed-boundary points instead. The transients of the adjoint simulations can be seen to correspond to several vortex-shedding periods before the two curves start to overlap.}\label{adjointforces}
\end{figure}

One way to visualize this time scale is to realize that the adjoint flow-field will be a transient simulation, even if based on a fully periodic limit-cycling base-flow. From a numerical point of view, given that the gradient of the controls is in general dependent on both the forward and adjoint state, the horizon should be long enough for its value at the start of the simulation not to be significantly affected by any further increase in the horizon length. Comparing the forces of the adjoint simulation from the first iteration of Run 4 to those of Run 6 (Fig.~\ref{adjointforces}) it is clear that the adjoint simulation experiences transients for longer than one vortex-shedding period before the two sets of forces overlap (recall that the adjoint simulation runs backwards in time). 

For the last few vortex-shedding periods of the forward simulation, the control will be therefore optimized according to what can be considered a ``short term'' control strategy, which may not be sustainable. In some cases, having several short horizons might just lead to a suboptimal but similar control waveform to the long horizon case, but in others, the two physical mechanisms leading to a reduction in the cost might be different altogether (as with the 10 convective time-unit horizons in Run 3). Note however that as the control can simply tend towards two different local minima, it is not a priori obvious which one will have the lowest cost (indeed, the drag is reduced more with fast constant rotation than with the oscillatory rotation in this case).

%%%

\section{Conclusions}\label{ccl}

In this article, optimal control is applied to the flow over a cylinder, where the total power, the mean-squared lift or the mean-squared drag were minimized using unconstrained rotation of the cylinder. The optimization horizons were chosen to be longer than in previous studies (where they typically are at most of the order of a vortex shedding period).

It was shown that at Reynolds numbers between 75 and 200, cylinder rotation can effectively suppress vortex shedding in the wake. The optimal control waveform was found to be phase-locked with vortex shedding, and the rotation is applied in a direction that weakens the vortices shed in the wake. The stabilizing action of the control on the wake is therefore similar to the high-amplitude and high-frequency open-loop optimum that is well documented in the literature, but feedback vastly reduces the required rotation rate.

It was found that this phase-locking behavior can only be achieved in a closed-loop control setting. Applying an almost identical but harmonic (i.e. open-loop) forcing signal instead results in a drift between the forcing and shedding phase, and eventually \textit{increases} drag. On the other hand, the optimal forcing decreases the drag by up to $19\%$ compared to the unforced case and the amplitude of the actuation required to keep the flow in this stabilized state approaches zero, with tangential velocities on the cylinder surface of the order of $0.1\%$ of the inflow velocity towards the end of the considered optimization window. This shows the importance of keeping the control waveform fully unconstrained, as previous studies where only periodic solutions were considered reached the high-amplitude and frequency \textit{open-loop} optimum rather than the stabilizing control obtained here.

The analysis of the influence of different optimization parameters confirmed the fact that the setup of the optimization -- including the choice of cost function and horizon length -- can strongly affect the final results. In particular, changing the horizon length can even change the local minimum that is identified and hence the mechanism through which the cost function is minimized. In this case, it was found an optimization window longer than the vortex shedding frequency was required in order for the converged optimal control to stabilize the wake in an efficient manner. It was found that the transients of the adjoint simulation can provide some important information regarding the adequacy of the chosen horizon length. If the horizon is long enough, the state of the adjoint solution and hence the control gradients will not be strongly affected by any further increase of the horizon length. On the other hand, if the horizon is too short, this can result in a ``short term'' optimization, whereby the reduction in the cost must be measured quickly but may be associated with a different control mechanism to the one corresponding to the ``long term'' optimum. In this case, with horizons of 10 convective time-units, the drag was minimized with a fast, nearly constant rotation of the cylinder, whereas with longer horizons of 50 convective time-units or more, the low amplitude, approximately zero-mean, oscillatory rotation of the cylinder described above was obtained.

In this article, we have demonstrated the importance of choosing a long enough optimization horizon when considering adjoint-based optimal control problems, especially when they are based on unsteady nonlinear base-flows. We have shown that increasing the horizon length not only leads to smoother waveforms but also to potentially large improvements in the results. Furthermore the smooth control waveforms make it possible to interpret the physical mechanisms that are responsible for the effectiveness of the control. Our work also confirmed that excessively long horizons can be detrimental to the convergence of the results and that changing the horizon length can result in a control waveform that uses an altogether different mechanism thus reaching a separate local minimum in the cost function.

%%%

\appendix

\section{Forward and adjoint solvers and procedures}\label{IBFS}

The immersed-boundary fractional step solver developed by Taira and Colonius\cite{Taira2007,colonius2008} solves the vorticity form of the incompressible Navier-Stokes equations (\ref{eq:nseq}) numerically. Further information regarding the solvers can also be found for instance in \cite{Ahuja2010} and \cite{JoeThesis}.

Using a uniform staggered Cartesian mesh in regions close to the cylinder, a Dirichlet boundary condition is imposed at a set of arbitrarily defined Lagrangian points, which define the body surface, as shown in Fig.~\ref{cylindergrid}. The spatially discretized form of the equation reads:
\begin{eqnarray}
\left\{
\begin{array}{l}
\dot\gamma+C^TE^T\tilde f = -\beta C^TC\gamma+C^T{\cal N}(q,\gamma)+bc,\\
ECs=\Delta\phi\hat u_t.
\end{array}\right.
\label{eq:discns}
\end{eqnarray}
Here, $\gamma$ represents the discretized circulation around any given cell (evaluated at the cell center) and $s$ is the discretized streamfunction (also evaluated at the cell center). 
$\hat u_t$ is a unit vector defined at each immersed-boundary point, which is oriented in the tangential direction to the body surface in the counterclockwise direction and $\phi$ is the time-varying control amplitude. $C$ is the discrete curl operator, constructed so that $q=Cs$, where ${q=\left[\begin{array}{cc}\!q_x^T&q_y^T\!\end{array}\right]^T}$ is the flux into the cell, whose $x$ and $y$ components $q_x$ and $q_y$ are evaluated at the left and bottom cell edges respectively. Similarly, we have $\gamma=C^Tq$, as well as ${s=(C^TC)^{-1}\gamma}$. The discrete Laplace operator is given by $-C^TC\gamma$. ${\tilde f=\left[\begin{array}{cc}\!\tilde f_x^T&\tilde f_y^T\!\end{array}\right]^T}$ are the discretized immersed forces with a scaling factor, evaluated at the immersed body points and $\tilde f_x$ and $\tilde f_y$ are the $x$ and $y$ components of the vector at each point. $E$ and $E^T$ are interpolation and regularization operators that allow evaluating the quantities defined on the Cartesian mesh at the immersed body point locations and vice-versa. ${\cal N}(q,\gamma)$ is the discretization of the nonlinear advection operator $u\times\omega$. Finally, $bc$ are the boundary conditions imposed at the edges of each domain and ${\beta=1/(\Delta^2Re)}$, where $\Delta$ is the uniform grid spacing and $Re$ is the Reynolds number.

In the present work, the body is moving horizontally with a velocity $U_\infty$ through the fluid. Staying in the \textit{body} reference frame, the (potential) flux of the incoming flow is therefore $-q_\infty$ and the total flux vector thus becomes $q-q_\infty$. Equations~(\ref{eq:discns}) therefore become:
\begin{eqnarray}
\left\{
\begin{array}{l}
\dot\gamma+C^TE^T\tilde f = -\beta C^TC\gamma+C^T{\cal N}(q-q_\infty,\gamma)+bc,\\
E(Cs-q_\infty)=\Delta\phi\hat u_t,
\end{array}\right.
\label{eq:discns2}
\end{eqnarray}
which we can write:
\begin{eqnarray}
\left\{
\begin{array}{l}
\dot\gamma={\cal F}(\gamma,\tilde f),\\
E(Cs-q_\infty)=\Delta\phi\hat u_t.
\end{array}\right.
\label{eq:discnsshort}
\end{eqnarray}

A second-order (explicit) Adam-Bashforth scheme is used to discretize the nonlinear term and the (implicit) Crank-Nicolson method is chosen for the linear terms. The time-stepping scheme is based on a diagonalization of the Laplacian using the fast Sine transform and a fractional step method, where the Poisson equation to be solved is only of the dimension of the number of immersed body force vector $\tilde f$.

In order to use far-field boundary conditions, an excessively large uniform Cartesian domain would be required for most open flows. Instead, the flow is solved on a series of nested uniform Cartesian grids, where each grid level is identical to the previous one, but has twice the physical extent and hence half the resolution, as illustrated in Fig.~\ref{gridlevs}. The coarsest grid is chosen to be large enough for the far-field boundary conditions to be justified there. Boundary conditions for all remaining grid levels can then be interpolated from the next grid level. The immersed-boundary points are only assumed to exist on the finest grid. At each time-step, the circulation is thus first obtained there, allowing the solution to be coarsified onto all other grid levels.

It is straightforward to evaluate the total force acting on the body in a given direction (and therefore the corresponding force coefficient) for a given flow state. This force can be obtained by summing the components of the forces at all the immersed-boundary points in the direction of interest. As a result, the three cost functions we consider in this article, introduced in Eq.~(\ref{eq:costP}, \ref{eq:costL}, \ref{eq:costD}), are calculated in the following way:

\begin{eqnarray}
{\cal J_P}&=&\int_0^T \Big\{ \tilde f\,^T Eq_\infty+ \tilde f\,^T \hat u_t \phi \Big\}dt,\nonumber\\
{\cal J_L}&=&\int_0^T \frac{1}{2} \tilde f_y\,^T \tilde f_y\,dt,\nonumber\\
{\cal J_D}&=&\int_0^T \frac{1}{2}\tilde f_x\,^T \tilde f_x\,dt\nonumber.
\end{eqnarray}

As shown in Appendix~\ref{adjointequations}, the discretized adjoint equations (\ref{eq:adjeqdisc}) are similar in form to the forward equations (\ref{eq:discns}), except for the advection terms and the fact that the adjoint equations run backwards in time from $t=T$ to $t=0$. A very similar solver to the one introduced here is therefore used for the adjoint equations, including the discretization and time-stepping schemes, the far-field boundary conditions, and the nested-grid algorithm.

%%%

\section{Adjoint Equations}\label{adjointequations}

Starting from the spatially discrete but time-continuous forward equations (\ref{eq:discnsshort}), a standard adjoint-based optimization procedure is used in this article. We first define a cost function that depends on the state variables $x$ and the imposed control $\phi$. In our case, it takes the form:
\begin{eqnarray}
{\cal J}(x,\phi)=\int_0^T{\cal I}(x,\phi)dt.
\end{eqnarray}
The state variable comprises of the circulation and the immersed-boundary forces ${x=\left[\begin{array}{cc}\!\gamma^T&\tilde f^T\!\end{array}\right]^T}$. The cost is augmented by Lagrange multipliers (the adjoint state) ${x^+=\left[\begin{array}{cc}\!\gamma^{+T}&\tilde f^{+T}\!\end{array}\right]^T}$ that constrain the state variables to respect the forward equations (\ref{eq:discnsshort}):
\begin{eqnarray}
\tilde{\cal J}(x,\phi,x^+)&=&\int_0^T\Big\{{\cal I}(x,\phi)-\gamma^{+T}\left(C^TC\right)^{-1}\left[ \dot\gamma-{\cal F}(x) \right]%\nonumber\\
-\tilde f^{+T}\left[ E(Cs-q_\infty)-\Delta\phi\hat u_t \right]\Big\}dt,\nonumber\\
&=&\int_0^T{\cal H}(x,\phi,x^+)dt.
\label{eq:Jaug}
\end{eqnarray}
Note that for the momentum equation, we use a non-trivial symmetric inner-product matrix $(C^TC)^{-1}$, so that $\langle\gamma,\gamma\rangle=\gamma^T(C^TC)^{-1}\gamma=q^Tq$, which is directly related to the kinetic energy of the flow. All variations of Eq.~({\ref{eq:Jaug}}) with respect to $x$, $x^+$ and $\phi$ must be zero for the solution to be optimal. Thus, setting variations with respect to $x^+$ to zero first, we obtain:
\begin{eqnarray}
\left\{
\begin{array}{l}
\dot \gamma={\cal F}(x),\\
E(Cs-q_\infty)=\Delta\phi\hat u_t.\nonumber
\end{array}
\right.
\end{eqnarray}
In other words, the forward equations must be enforced throughout the horizon. Next, setting variations with respect to $x$ to zero can be shown to result in the following \textit{adjoint equations}:
\begin{eqnarray}
\left\{
\begin{array}{l}
-\dot\gamma^++C^TE^T\tilde f^+=-\beta C^TC\gamma^+ + (C^TC)\left(\frac{\partial {\cal N}}{\partial \gamma}\right)^Tq^+ + (C^TC)\left(\frac{\partial {\cal I}}{\partial \gamma}\right)^T+bc^+,\\
%-\dot\gamma^+ + \left(\frac{\partial {\cal F}}{\partial \gamma}\right)^T&=&\left(\frac{\partial {\cal I}}{\partial \gamma}\right)^T\nonumber\\
%
ECs^+=\left(\frac{\partial {\cal I}}{\partial \tilde f}\right)^T,\\
\gamma^+(T)=(C^TC)\left(\frac{\partial {\cal I(T)}}{\partial \gamma}\right)^T,
\end{array}
\right.
\label{eq:adjeqdisc}
\end{eqnarray}
where $q^+=Cs^+$ are the adjoint fluxes, $s^+=(C^TC)^{-1}\gamma^+$ is the adjoint streamfunction, and $bc^+$ are the adjoint boundary conditions. All quantities here are defined in an analogous manner to their counterpart in the forward equations. Equations~(\ref{eq:adjeqdisc}) must also be satisfied throughout the horizon, but they are integrated backwards in time from the initial condition at $t=T$ (given by the third equation in (\ref{eq:adjeqdisc})). For the cost functions considered here, we have simply $\gamma^+(T)=0$, since we do not have a cost on the final flow condition. In fact, it can be shown that these equations can also be obtained by directly discretizing Eq.~(\ref{eq:adjeq}), which are their continuous equivalent.

Note that the adjoint equations are very similar to the forward equations (\ref{eq:discns}), except for the negative time derivative and the advection term. This term can be shown to be a discretization of $\nabla\times\left(\omega\times u^+\right)-\nabla^2\left(u^+\times u\right)$ and therefore depends on the (unsteady, nonlinear) state of the forward simulation at time $t$ (since $u$ and $\omega$ are needed to compute it). The entire forward base-flow $\gamma(t)$ for all $t$ such that $0\le t\le T$ is thus required to compute the adjoint solution.

Note also that the second equation in (\ref{eq:adjeqdisc}) sets the ``slip flux'': $ECs^+=q_{slip}^+=\left(\partial {\cal I}/\partial \tilde f\right)^T$ at the immersed-boundary points and is dependent on the cost function. For the three cost functions considered in this article -- i.e. ${\cal J_P}$, ${\cal J_L}$, and ${\cal J_D}$, given in Eq.~(\ref{eq:costP},\ref{eq:costL},\ref{eq:costD}) -- the slip flux is given respectively by:
\begin{eqnarray}
q_{slip,{\cal P}}^+&=&\Delta\phi\hat u_t+\Delta Eq_\infty,\nonumber\\
q_{slip,{\cal L}}^+&=&\left[\begin{array}{cc}\!0&\tilde f_y^T\!\end{array}\right]^T,\nonumber\\
q_{slip,{\cal D}}^+&=&\left[\begin{array}{cc}\!\tilde f_x^T& 0\!\end{array}\right]^T,\nonumber
\end{eqnarray}
The right-hand-side forcing term of the adjoint momentum equation (the first equation in (\ref{eq:adjeqdisc})) is also dependent on the cost in general: for the cost functions considered in this article we have simply $\partial {\cal I}/\partial \gamma=0$.

Finally, taking variations with respect to $\phi$ yields the control gradient:
\begin{eqnarray}
{\cal G}=\left(\frac{\partial {\cal H}}{\partial \phi}\right)^T=\left(\frac{\partial {\cal I}}{\partial \phi}\right)^T+\Delta\tilde f^{+T}\hat u_t.\nonumber
\end{eqnarray}
The gradient will only be zero if the control is exactly optimal, which is not the case in general. Its direction is therefore used to update the current control guess in a locally optimal manner. For the cost functions considered here -- again ${\cal J_P}$, ${\cal J_L}$, and ${\cal J_D}$, given in Eq.~(\ref{eq:costP},\ref{eq:costL},\ref{eq:costD}) -- the expression of the three gradients are respectively:
\begin{eqnarray}
{\cal G_P}&=&\Delta\left(\tilde f^{\: T}+\tilde f^{+T}\right)\hat u_t,\nonumber\\
{\cal G_L}&=&\Delta\tilde f^{+T}\hat u_t,\nonumber\\
{\cal G_D}&=&\Delta\tilde f^{+T}\hat u_t.\nonumber
\end{eqnarray}

%%%

\end{document}